\documentclass[journal]{IEEEtran}
\usepackage{amssymb}
\usepackage{color}
\usepackage{multirow}
\usepackage{mathrsfs}
\newtheorem{theorem}{Theorem}
\newtheorem{lemma}{Lemma}
\newtheorem{proposition}{Proposition}

\newtheorem{definition}{Definition}

\usepackage{cite}
\ifCLASSINFOpdf
  \usepackage[pdftex]{graphicx}
  \graphicspath{{../pdf/}{../jpeg/}}
  \DeclareGraphicsExtensions{.pdf,.jpeg,.png}
\else
  \usepackage[dvips]{graphicx}
  \graphicspath{{../eps/}}
  \DeclareGraphicsExtensions{.eps}
\fi
\usepackage[cmex10]{amsmath}
\usepackage{array}
\usepackage[tight,footnotesize]{subfigure}
\usepackage{stfloats}

\usepackage{arydshln}
\usepackage{rotating}
\usepackage{microtype}
\IEEEoverridecommandlockouts

\usepackage{epstopdf}
\begin{document}
\title{The DoF Region of the Three-Receiver MIMO Broadcast Channel with Side Information and Its Relation to Index Coding Capacity}

\author{\IEEEauthorblockN{Behzad Asadi, Lawrence Ong, and Sarah J.\ Johnson}
\thanks{The authors are with the School of Electrical Engineering and Computer Science, The University of Newcastle, Australia (e-mail: behzad.asadi@uon.edu.au, lawrence.ong@cantab.net, sarah.johnson@newcastle.edu.au). This work is supported by the Australian Research Council under grants FT110100195, FT140100219, and DP150100903.}
}

\maketitle

\begin{abstract}
We consider the three-receiver Gaussian multiple-input multiple-output (MIMO) broadcast channel with an arbitrary number of antennas at each of the transmitter and the receivers. We investigate the degrees-of-freedom (DoF) region of the channel when each receiver requests a private message, and may know some of the messages requested by the other receivers as receiver message side information (RMSI). We establish the DoF region of the channel for all 16 possible non-isomorphic RMSI configurations by deriving tight inner and outer bounds on the region. To derive the inner bounds, we first propose a scheme for each RMSI configuration which exploits both the null space and the side information of the receivers. We then use these schemes in conjunction with time sharing for 15 RMSI configurations, and with time sharing and two-symbol extension for the remaining one. To derive the outer bounds, we construct enhanced versions of the channel for each RMSI configuration, and upper bound their DoF region. After establishing the DoF region, in the case where all the nodes have the same number of antennas, we introduce some common properties of the DoF region, and the capacity region of the index coding problem.
\end{abstract}
\IEEEpeerreviewmaketitle

\section{Introduction}
Deploying multiple-antenna nodes in communication networks offers several significant benefits including a multiplicative increase in transmission rates~\cite{MIMO}. This makes multiple-antenna nodes an integral part of data communication networks.

Broadcast channels~\cite{BC}, as a building block of such networks, model the scenario where a transmitter wants to send a number of messages to multiple receivers through a shared medium. The capacity region of the Gaussian multiple-input multiple-output (MIMO) broadcast channel with two receivers is know when the receivers have both common- and private-message requests~\cite{MIMOBCwithCommon}; the capacity region of the channel with more than two receivers is also known when the receivers have only private-message requests~\cite{MIMOBCwithoutCommon}. These capacity results quantify the increase in transmission rates achieved by increasing the number of antennas at the nodes.

In communication networks, receivers may know a priori some of the messages requested by other receivers as receiver messages side information (RMSI). This form of side information
appears in, for example, multimedia broadcasting with packet loss, and the downlink phase of applications
modeled by multi-way relay channels~\cite{MWRCFullExchange}. It is known that using RMSI in code design can increase transmission rates over the Gaussian broadcast channel with single-antenna nodes~\cite{BCwithSI2UsersGeneral,ThreeReceiverAWGNwithMSI}. The capacity region of the Gaussian MIMO broadcast channel with two receivers is known when each receiver knows the private message requested by the other receiver as RMSI, i.e., complementary RMSI~\cite{MIMOandRMSI}. For the considered setting (which is equivalent to multicasting a common message to the receivers), this result quantifies the further possible increase in transmission rates due to RMSI. However, the capacity region of the Gaussian MIMO broadcast channel with non-complementary RMSI is not known. It is particularly difficult to characterize the capacity region where there are more than two receivers. This is because non-complementary RMSI leads to a scenario that implicitly involves transmitting both common and private messages, and as mentioned earlier, the capacity region of the Gaussian MIMO broadcast channel with more than two receivers has resisted solution when the receivers have both common- and private-message requests.

\begin{figure}[t]
	\centering
	\includegraphics[width=0.45\textwidth]{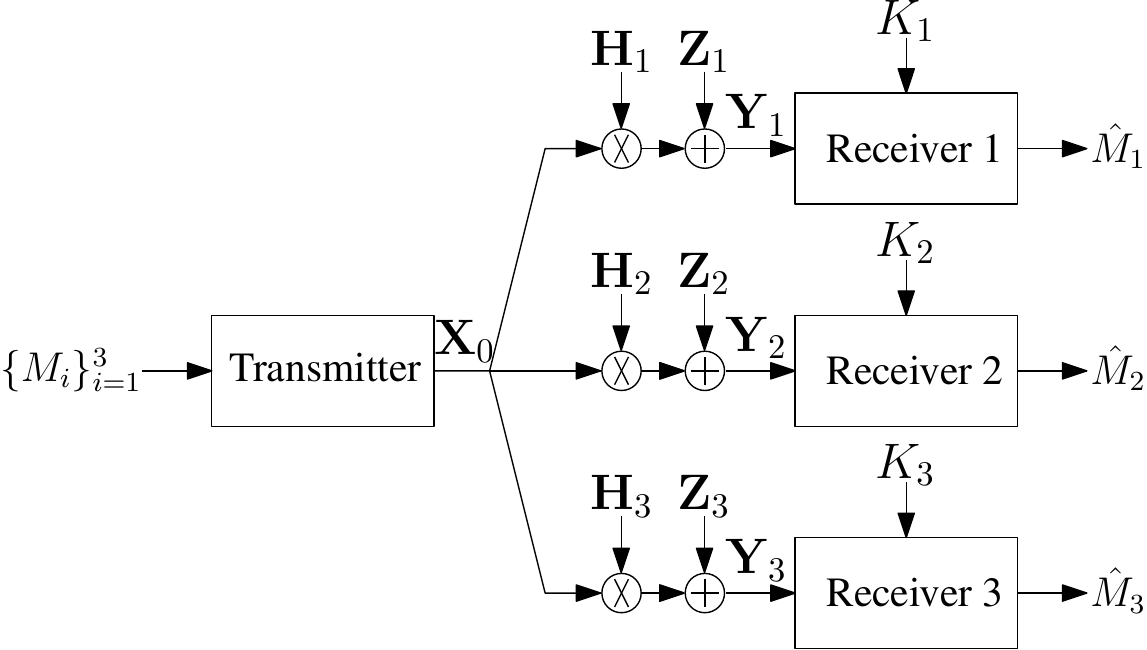}
	\vskip-0pt
	\caption{The three-receiver Gaussian MIMO broadcast channel where $\mathbf{X}_{0}\in\mathbb{C}^{N_0\times 1}$, $\mathbf{H}_i\in\mathbb{C}^{N_i\times N_0}$, $\mathbf{Y}_i\in\mathbb{C}^{N_i\times 1}$, $\mathbf{Z}_i\in\mathbb{C}^{N_i\times 1}$, and $\mathbf{Z}_i\sim\mathcal{CN}(\mathbf{0},\mathbf{I}_i)$, $i\in\{1,2,3\}$. $M_i$ is the message requested by receiver~$i$, $K_i\subseteq \{M_1,M_2,M_3\}\setminus M_i$ is the set of messages known a priori to receiver~$i$, and $\hat{M}_i$ is the decoded message at receiver~$i$.} 
	\vskip-0pt
	\label{Fig:SystemModel}
\end{figure}

When the capacity region of a Gaussian channel is difficult to determine, it has been of great interest to derive the degrees-of-freedom (DoF) region of the channel, such as for the MIMO interference channel~\cite{DoF2PairIC}, and the MIMO X channel~\cite{DoF2PairXJafar,DoF2PairXMoh}. The DoF region characterizes the limit of the capacity region normalized by the logarithm of the transmission power as the power goes to infinity. So establishing the DoF region of the Gaussian MIMO broadcast channel with RMSI quantifies the further possible increase in transmission rates (due to RMSI) in the high signal-to-noise ratio regime. Concerning DoF results for MIMO broadcast channels with RMSI, Jafar and Shamai~\cite{DoF2PairXJafar} characterized the sum-DoF of the Gaussian MIMO X channel (which includes the two-receiver Gaussian MIMO broadcast channel as a special case) where all the nodes have the same number of antennas, and one of the receivers knows a priori one of the messages requested by the other receiver. Zhang and Elia~\cite{MISOBCReceiverCache} considered the fading broadcast channel with an $N$-antenna transmitter, and $N$ single-antenna receivers equipped with a cache. They assumed that the transmitter knows partially the current channel state, and perfectly a delayed version of it. They characterized the sum-DoF within a multiplicative factor of four.

\subsection{Contributions}\label{Sec:mainresults}
In this work, we consider the three-receiver Gaussian MIMO broadcast channel with an arbitrary number of antennas at the transmitter and the receivers. We assume that (i) channel matrices are known to the transmitter and all the receivers, (ii) the receivers have private-message requests, and (iii) each receiver may know some of the messages requested by the other receivers as RMSI. This results in 16 possible non-isomorphic RMSI configurations in the sense that we cannot transform one to another by re-labeling the receivers and their requested messages. We derive tight inner and outer bounds on the DoF region of the channel for all 16 possible RMSI configurations, thereby establishing their DoF region. We construct our proposed schemes by utilizing both the null space and the side information of the receivers. We derive our outer bounds by upper bounding the DoF region for enhanced versions of the channel. In addition, in the case where all the nodes have the same number of antennas, we draw an analogy between the DoF region, and the capacity region of the index coding problem~\cite{CapacityRegionIndexCoding1}.

\section{System Model}\label{Section:SystemModel}
We consider the three-receiver Gaussian MIMO broadcast channel, depicted in Fig.~\ref{Fig:SystemModel}, where the transmitter is equipped with $N_0$ antennas, and receiver~$i$, $i\in\{1,2,3\}$, is equipped with $N_i$ antennas. In this channel, at time instant~$j$, we have
\begin{align*}
\mathbf{Y}_{i,j}=\mathbf{H}_{i}\mathbf{X}_{0,j}+\mathbf{Z}_{i,j}.
\end{align*}
$\mathbf{X}_{0,j}\in\mathbb{C}^{N_0\times 1}$ is the transmitted vector where $\mathbb{C}$ represents the set of complex numbers, $\mathbf{Y}_{i,j}\in\mathbb{C}^{N_i\times 1}$ is the channel-output vector at receiver~$i$, $\mathbf{H}_i\in\mathbb{C}^{N_i\times N_0}$ is the channel matrix between the transmitter and receiver~$i$, and $\mathbf{Z}_{i,j}\in\mathbb{C}^{N_i\times 1}$ is a white circularly symmetric complex Gaussian noise with zero mean and an $N_i\times N_i$ identity covariance matrix, i.e., $\mathbf{Z}_{i,j}\sim\mathcal{CN}(\mathbf{0},\mathbf{I}_i)$. We represent random variables using upper-case letters, and their realizations using the corresponding lower-case letters. We denote the $k$-th entry of a column vector $\mathbf{A}$ as $A_{[k]}$, the entry in the $k$-th row and the $\ell$-th column of a matrix $\mathbf{B}$ as $B_{[k\ell]}$, the $k$-th row of a matrix $\mathbf{B}$ as $\mathbf{B}_{[k:]}$, and the $\ell$-th column of a matrix $\mathbf{B}$ as $\mathbf{B}_{[:\ell]}$. Then we have
\begin{align*}
Y_{i,j[k]}=\mathbf{H}_{i[k:]}\mathbf{X}_{0,j}+Z_{i,j[k]},\;\; k\in\{1,2,\ldots,N_i\}.
\end{align*}
We assume that the channel coefficients, $H_{i[kl]}$, are generated independently according to a continuous distribution. This yields the rank of the matrix $\mathbf{H}_i,\;i\in\{1,2,3\} ,$ to be almost surely $\min\{N_i,N_0\}$, i.e., full rank. This also yields the rank of the matrix 
\begin{align*}
\begin{bmatrix}
\mathbf{H}_1^T&\mathbf{H}_2^T&\mathbf{H}_3^T
\end{bmatrix}
\end{align*}
to be almost surely $\min\{N_0,\sum_{i=1}^{3}N_i\}$ where $[\cdot]^T$ denotes the transpose operation. 

Considering $n$ uses of the channel, the intended message for receiver~$i$, $M_i,\;i\in\{1,2,3\},$ is an $nR_i$-bit message, and is uniformly distributed over the set $\mathcal{M}_i=\{0,1,\ldots,2^{nR_i}-1\}$. The transmitted codeword $\mathbf{X}_{0}^n=\left(\mathbf{X}_{0,1},\mathbf{X}_{0,2},\ldots,\mathbf{X}_{0,n}\right)$, which is a function of source messages, $\{M_i\}_{i=1}^{3}$, has the power constraint of 
\begin{equation}\label{powerconstraint}
\sum_{j=1}^{n}\text{tr}\left(\mathbf{X}_{0,j}(m_1,m_2,m_3)\mathbf{X}_{0,j}^*(m_1,m_2,m_3)\right)\leq nP,
\end{equation}
for any $(m_1,m_2,m_3)$ where $[\cdot]^*$ denotes the conjugate transpose operation, and $\text{tr}(\cdot)$ the trace of a square matrix.

\begin{figure}[t]
	\centering
	\includegraphics[width=0.45\textwidth]{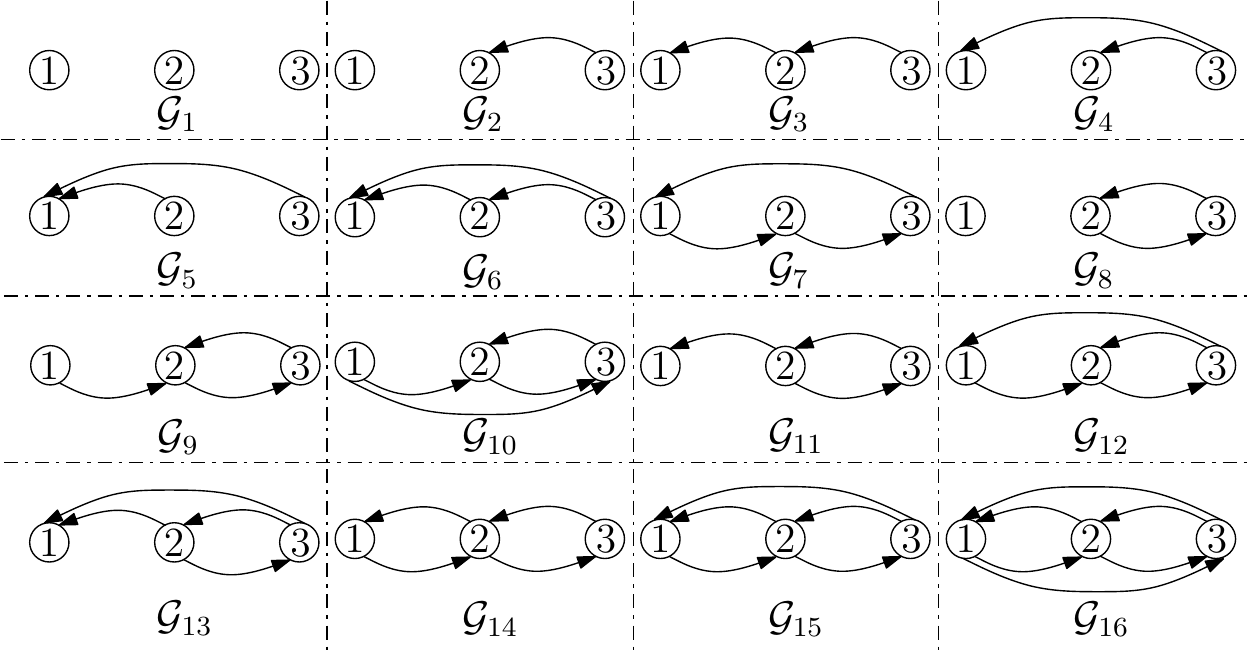}
	\vskip-0pt
	\caption{Non-isomorphic side information graphs modeling all possible side information configurations.} 
	\vskip-0pt
	\label{Fig:Graphs}
\end{figure}

Receiver~$i$, $i\in\{1,2,3\}$, knows a priori an ordered set of messages $K_i\subseteq\{M_1,M_2,M_3\}\setminus M_i$ as RMSI. We model the side information configuration of the channel by a side information graph $\mathcal{G}=(\mathcal{V}_\mathcal{G}, \mathcal{A}_\mathcal{G})$ where $\mathcal{V}_\mathcal{G}=\{1,2,3\}$ is the set of \textit{vertices}, and $\mathcal{A}_\mathcal{G}$ is the set of \textit{arcs}. An arc from vertex~$i$ to vertex~$i'$ exists if and only if receiver~$i$ knows $M_{i'}$ as side information. Then the set of outneighbors of vertex~$i$ is $\mathcal{O}_i=\{i'\mid M_{i'}\in K_i\}$. As an example, the graph
\begin{center}
\vskip-3pt
\includegraphics[width=0.11\textwidth]{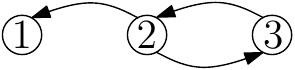}
\end{center}
\vskip-7pt
represents the case where $K_1=\emptyset$, $K_2=\{M_1,M_3\}$, and $K_3=\{M_2\}$. Any side information configuration can be modeled by one of the 16 graphs shown in Fig.~\ref{Fig:Graphs}. These are all possible non-isomorphic RMSI configurations in the sense that we cannot transform one to another by re-labeling the receivers and their requested messages. For instance, the graph
\begin{center}
	\vskip-3pt
	\includegraphics[width=0.11\textwidth]{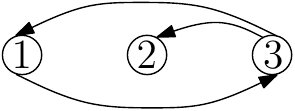}
\end{center}
\vskip-7pt
is transformed to $\mathcal{G}_{11}$ by re-labeling  vertex~1 as vertex~3, vertex~2 as vertex~1, and vertex~3 as vertex~2.

A $(2^{nR_1},2^{nR_2},2^{nR_3},n)$ code for the channel consists of an encoding function
\begin{align*}
f: \mathcal{M}_1\times\mathcal{M}_2\times\mathcal{M}_3\rightarrow \mathbb{C}^{N_0\times n},
\end{align*}
with the power constraint in \eqref{powerconstraint} where $\times$ denotes the Cartesian product when it is used for sets. Then the transmitted codeword is $\mathbf{X}_0^n=f(M_1,M_2,M_3)$. This code also consists of decoding functions 
\begin{align*}
g_i:\mathbb{C}^{N_i\times n}\times\mathcal{K}_i\rightarrow \mathcal{M}_i,\;i\in\mathcal{V}_\mathcal{G},
\end{align*}
where
\begin{align*}
\mathcal{K}_i=\bigotimes_{\ell\in\mathcal{O}_i}\mathcal{M}_\ell.
\end{align*}
For instance, if $K_1=\{M_2,M_3\}$, we have $\mathcal{K}_1=\mathcal{M}_2\times\mathcal{M}_3$. Then the decoded message at receiver~$i$ is $\hat{M}_i=g_i\left(\mathbf{Y}_i^{n},K_i\right)$. The average probability of error for this code is defined as 
\begin{align*}
P_e^{(n)}=P\left((\hat{M}_1,\hat{M}_2,\hat{M}_3)\neq({M}_1,{M}_2,{M}_3)\right).
\end{align*}
\begin{definition}
	A rate triple $(R_1(P),R_2(P),R_3(P))$ is said to be achievable if there exists a sequence of $(2^{nR_1},2^{nR_2},2^{nR_3},n)$ codes with $P_e^{(n)}\rightarrow 0$ as $n\rightarrow \infty$.
\end{definition}

\begin{definition}
	The capacity region of the channel, $\mathcal{C}(P)$, is the closure of the set of all achievable rate triples $(R_1(P),R_2(P),R_3(P))$.
\end{definition}

\begin{definition}
	A DoF triple $(d_1,d_2,d_3)$ is said to be achievable if, for any $(w_1,w_2,w_3)\in\mathbb{R}^3_{+}$, there exists an achievable rate triple $(R_1(P),R_2(P),R_3(P))$ such that
	\begin{align*}
		\sum_{i=1}^3 w_id_i\leq\underset{P\rightarrow \infty}{\lim\sup}\left[\sum_{i=1}^{3}\frac{w_iR_i(P)}{\log{P}}\right], 
	\end{align*}
	where $\mathbb{R}_{+}$ represents the set of positive real numbers.
\end{definition}

\begin{definition}
	The DoF region of the channel is the set $\mathcal{D}$ which is defined as~\cite{DoF2PairXJafar}
	\begin{align*}
	&\mathcal{D}\hskip-3pt=\hskip-3pt\Bigg{\{}\hskip-2pt(d_1,d_2,d_3)\in\mathbb{R}^3_+\mid \forall (w_1,w_2,w_3)\in\mathbb{R}^3_+ \\
	&\hskip33pt\sum_{i=1}^3 w_id_i\leq\underset{P\rightarrow \infty}{\lim\sup}\left[\left[\underset{\mathcal{C}(P)}{\sup}{\sum_{i=1}^{3}w_iR_i(P)}\right]\frac{1}{\log{P}}\right]\Bigg{\}}.
	\end{align*}
\end{definition}

\section{DoF Region with RMSI}
In this section, we characterize the DoF region of the channel for all 16 possible non-isomorphic side information configurations, stated as Theorem~\ref{Theorem:DoF}.

\begin{theorem}\label{Theorem:DoF}
	The DoF region of the three-receiver Gaussian MIMO broadcast channel with the side information graph $\mathcal{G}_k$ is $\mathcal{D}_k$ where
\begin{align*}
\mathcal{D}_k=\Big{\{}(d_1,d_2,d_3)\in&\mathbb{R}_+^3 \mid\\ 
d_1+d_2+d_3&\leq N_0, \\
d_i&\leq N_i,\;i\in\mathcal{V}_\mathcal{G}\Big{\}},\;k\in\{1,2,\ldots, 6\},\\
\mathcal{D}_7=\Big{\{}(d_1,d_2,d_3)\in&\mathbb{R}_+^3 \mid\\
d_1+d_2&\leq N_0,\\
d_1+d_3&\leq N_0,\\
d_2+d_3&\leq N_0,\\
d_i&\leq N_i,\;i\in\mathcal{V}_\mathcal{G}\Big{\}},
\end{align*}
\begin{align*}
\mathcal{D}_k=\Big{\{}(d_1,d_2,d_3)\in&\mathbb{R}_+^3 \mid\\
d_1+d_2&\leq N_0,\\
d_1+d_3&\leq N_0,\\
d_1+d_2+d_3&\leq\max\{N_0,N_2+N_3\},\\
d_i&\leq N_i,\;i\in\mathcal{V}_\mathcal{G}\Big{\}},\;k\in\{8,9,10\},\\
\mathcal{D}_k=\Big{\{}(d_1,d_2,d_3)\in&\mathbb{R}_+^3 \mid\\
d_1+d_2&\leq N_0,\\
d_1+d_3&\leq N_0,\\
d_i&\leq N_i,\;i\in\mathcal{V}_\mathcal{G}\Big{\}},\;k\in\{11,12,13\},\\
\mathcal{D}_k=\Big{\{}(d_1,d_2,d_3)\in&\mathbb{R}_+^3 \mid\\
d_1+d_3&\leq N_0,\\
d_2&\leq N_0,\\
d_i&\leq N_i,\;i\in\mathcal{V}_\mathcal{G}\Big{\}},\;k\in\{14,15\},\\
\text{and}\hskip90pt&\\
\mathcal{D}_{16}=\Big{\{}(d_1,d_2,d_3)\in&\mathbb{R}_+^3 \mid
d_i\leq\min\{N_0,N_i\},\;i\in\mathcal{V}_\mathcal{G}\Big{\}}.
\end{align*}
\end{theorem}
\begin{IEEEproof}
We prove the achievability of the integer DoF points within $\mathcal{D}_k,\;k\in\{1,2,\ldots,16\}$, in Section~\ref{Sec:IntegerDoF}, the achievability of the fractional DoF points in Section~\ref{Sec:FractionalDoF}, and the converse in Section~\ref{Sec:ConverseDoF}.
\end{IEEEproof}
\section{Achieving Integer DoF Points}\label{Sec:IntegerDoF}
In this section, we prove the achievability of all the integer points $(d_1,d_2,d_3)\in\mathbb{Z}_{+}^3\cap \mathcal{D}_k$, $k\in\{1,2,\ldots,16\}$, for the channel with $\mathcal{G}=\mathcal{G}_k$, where $\mathbb{Z}_+$ represents the set of positive integer numbers. 

It is well-known that all the integer points within the region $\mathcal{D}_1$ are achievable for the three-receiver Gaussian MIMO broadcast channel without RMSI~\cite{DoFNoCSIT2}. Then these points are also achievable for the channel with $\mathcal{G}=\mathcal{G}_k,\;k\in\{2,\ldots,6\}$ as the receivers have some extra side information. Hence, we just need to present the proof for the channel with $\mathcal{G}=\mathcal{G}_k,\;k\in\{7,8,\ldots,16\}$. 

To construct the transmission scheme, we first construct three subcodebooks. Subcodebook~$i$, $i\in\mathcal{V}_\mathcal{G}$, consists of $2^{nR_i}$ i.i.d. codewords 
\begin{align*}
\mathbf{X}^n_i(m_i)=(\mathbf{X}_{i,1}(m_i),\mathbf{X}_{i,2}(m_i),\ldots,\mathbf{X}_{i,n}(m_i)),
\end{align*}
generated according to $\prod_{j=1}^{n}p_{{\mathbf{X}_i}}(\mathbf{x}_{i,j})$ where $\mathbf{X}_{i,j}\in\mathbb{C}^{d_i\times 1}$, $\mathbf{X}_i\sim\mathcal{CN}(\mathbf{0},\mathbf{\Sigma}_i)$, $\mathbf{\Sigma}_i$ is a $d_i\times d_i$ diagonal matrix, $\text{tr}(\mathbf{\Sigma}_i)=P_i$, and $\sum_{i=1}^{3}P_i=P$. We then construct the transmitted codeword as
\begin{align*}
\mathbf{X}_{0,j}(m_1,m_2,m_3)=\sum_{i=1}^{3}\mathbf{V}_{i}\mathbf{X}_{i,j}(m_i),\;j\in\{1,2,\ldots,n\},
\end{align*}
where $\mathbf{V}_{i}\in\mathbb{C}^{N_0\times d_i}$ has columns, each of Euclidean norm one. Using this scheme, $(d_1,d_2,d_3)$ is achievable if we can choose the matrices $\{\mathbf{V}_{i}\}_{i=1}^{3}$ such that, at receiver~$i$, $i\in\mathcal{V}_\mathcal{G}$, we have  
\begin{align}
\mathbf{H}_i\mathbf{V}_{i[:\ell]}\notin\text{span}\left(\mathcal{F}_{i}\setminus \mathbf{H}_i\mathbf{V}_{i[:\ell]}\right),\;\ell\in\{1,2,\ldots,d_i\},\label{MainAchievCondition}
\end{align}
where $\mathcal{F}_i$ is the set of column vectors
\begin{align*}
\mathcal{F}_{i}=\Big{\{}\mathbf{H}_{i}\mathbf{V}_{i'[:\ell']}\mid i'\in\mathcal{V}_\mathcal{G}\hskip-3pt\setminus\hskip-3pt\mathcal{O}_i,\; \ell'\in\{1,2,\ldots,d_{i'}\}\Big{\}}.
\end{align*}
This is because, at receiver~$i$, we then can always find a vector $\mathbf{\Phi}_{i\ell}\in\mathbb{C}^{N_i\times 1}$, $\ell\in\{1,2,\ldots,d_i\}$, which is orthogonal to all the vectors in $\mathcal{F}_{i}\setminus \mathbf{H}_i\mathbf{V}_{i[:\ell]}$ but not to $\mathbf{H}_i\mathbf{V}_{i[:\ell]}$. This, i.e., $\mathbf{\Phi}^T_{i\ell}\mathbf{Y}_i^n$ for each $\ell$, provides us with an interference-free space dimension at receiver~$i$ which is equivalent to the output of a Gaussian single-antenna point-to-point channel.

The columns of the matrices $\{\mathbf{V}_{i}\}_{i=1}^{3}$ are generated either randomly according to an isotropic distribution, or using zero forcing~\cite{MIMOBook}. In this work, it suffices to consider zero forcing at individual receivers, and simultaneously at receivers~2 and~3. Then zero-forcing columns are selected from the columns of the matrices $\mathbf{S}_i\in\mathbb{C}^{N_0\times r_i},\;i\in\mathcal{V}_\mathcal{G}$, and $\mathbf{S}_{23}\in\mathbb{C}^{N_0\times r_{23}}$ where $r_i=(N_0-N_i)^+$, $r_{23}=(N_0-N_2-N_3)^+$, and $(a)^+=\max\{0,a\}$. The matrix $\mathbf{S}_i,\;i\in\mathcal{V}_\mathcal{G}$, is randomly generated in the null space of $\mathbf{H}_i$, i.e., $\mathbf{H}_i\mathbf{S}_{i}=\mathbf{0}$. The matrix $\mathbf{S}_{23}$ is randomly generated in the intersection of the null spaces of $\mathbf{H}_2$ and $\mathbf{H}_3$, i.e.,
\begin{align*}
\begin{bmatrix}
\mathbf{H}_2\\
\mathbf{H}_3
\end{bmatrix}\mathbf{S}_{23}=\mathbf{0}.
\end{align*}
Note that the rank of the matrix $[\mathbf{S}_{i},\mathbf{S}_{23}]\in\mathbb{C}^{N_0\times (r_i+r_{23})},\;i\in\{2,3\}$, is then almost surely $r_i$.

In the rest of this section, we show how the matrices $\{\mathbf{V}_{i}\}_{i=1}^{3}$ are chosen for the channel with $\mathcal{G}=\mathcal{G}_k,\;k\in\{7,8,\ldots,16\}$ in order to achieve all the integer points within the region $\mathcal{D}_k$.

\underbar{$\mathcal{G}=\mathcal{G}_7$:}
We choose the first $\min\{r_q,d_i\}$ columns of $\mathbf{V}_{i}$, $i\in\mathcal{V}_\mathcal{G}$, from the columns of $\mathbf{S}_q$ where $q=(i\hskip-4pt\mod 3)+1$, and we randomly generate the remaining $(d_i-r_q)^+$ columns. 

Using these matrices, we can almost surely have $d_1$ interference-free dimensions at receiver~1 if 
\begin{align*}
d_1+(d_3-r_1)^+&\leq \min\{N_0,N_1\}.
\end{align*}
This is because this condition yields the non-zero columns of the matrix $\left[\mathbf{H}_1\mathbf{V}_1,\mathbf{H}_1\mathbf{V}_3\right]$ to be almost surely linearly independent. Consequently, as receiver~1 knows $M_2$ as RMSI, condition~\eqref{MainAchievCondition} is satisfied at this receiver.

Similarly, we can almost surely have $d_2$ interference-free dimensions at receiver~2, and $d_3$ interference-free dimensions at receiver~3 if
\begin{align*}
d_2+(d_1-r_2)^+&\leq \min\{N_0,N_2\},\\
d_3+(d_2-r_3)^+&\leq \min\{N_0,N_3\}.
\end{align*}
Considering that $r_i+\min\{N_0,N_i\}=N_0,\;i\in\mathcal{V}_\mathcal{G}$, this completes the achievability proof of all the positive integer points within $\mathcal{D}_7$.

\underbar{$\mathcal{G}=\mathcal{G}_k,\;\;k\in\{8,9,10\}$:}  
We just need to prove achievability for the channel with $\mathcal{G}=\mathcal{G}_8$ as each $K_i$ for $\mathcal{G}_8$ is a subset of the corresponding one for $\mathcal{G}_k,\;k\in\{9,10\}$.

To construct $\mathbf{V}_1$, we select the first $\min\{r_{23},d_1\}$ columns of $\mathbf{V}_1$ from the columns of $\mathbf{S}_{23}$. For the remaining $(d_1-r_{23})^+$ columns, we select $r'_2$ columns from $\mathbf{S}_2$, $r'_3$ columns from $\mathbf{S}_3$, and randomly generate $(d_1-r_{23})^+-r'_2-r'_3$ columns where
\begin{align}
r'_2&\leq r_2-r_{23},\label{G8Cond1}\\
r'_3&\leq r_3-r_{23},\label{G8Cond2}\\
r'_2+r'_3&\leq (d_1-r_{23})^+.\label{G8Cond3}
\end{align}
Conditions~\eqref{G8Cond1} and~\eqref{G8Cond2} are imposed to ensure that $\mathbf{V}_1$ is almost surely full column rank.

To construct $\mathbf{V}_2$, and $\mathbf{V}_3$, we define $i_\text{max}$ as an arbitrary element of the set $\{2,3\}$ at which $d_{i_\text{max}}=\max\{d_2,d_3\}$, and $i_\text{min}$ as the other element of this set. We choose the first $\min\{r_1,d_{i_\text{max}}\}$ columns of $\mathbf{V}_{i_\text{max}}$ from the columns of $\mathbf{S}_1$, and randomly generate the remaining $(d_{i_\text{max}}-r_1)^+$ columns. We then choose the columns of $\mathbf{V}_{i_\text{min}}$ to be the same as the first $d_{i_\text{min}}$ columns of $\mathbf{V}_{i_\text{max}}$.

Using these matrices, we can almost surely have $d_1$ interference-free dimensions at receiver~1 if
\begin{align}
d_1+(\max\{d_2,d_3\}-r_1)^+&\leq \min\{N_0,N_1\}.\label{G8Cond4}
\end{align}
This is because, if condition~\eqref{G8Cond4} holds, the non-zero columns of the matrix $[\mathbf{H}_1\mathbf{V}_1,\mathbf{H}_1\mathbf{V}_{i_\text{max}}]$ are almost surely linearly independent. Consequently, condition~\eqref{MainAchievCondition} is satisfied at this receiver.

As receiver~2 knows $M_3$ as RMSI, also, we can almost surely have $d_2$ interference-free dimensions at receiver~2 if
\begin{align}
d_2+(d_1-r_{23})^+-r'_2&\leq \min\{N_0,N_2\},\label{G8Cond5}
\end{align}
and, as receiver~3 knows $M_2$ as RMSI, we can almost surely have $d_3$ interference-free dimensions at receiver~3 if
\begin{align}
d_3+(d_1-r_{23})^+-r'_3&\leq \min\{N_0,N_3\}.\label{G8Cond6}
\end{align}

Applying Fourier-Motzkin method to conditions~\eqref{G8Cond1}--\eqref{G8Cond6} in order to  eliminate $r'_2$ and $r'_3$ makes the proof complete for these configurations.
 
\underbar{$\mathcal{G}=\mathcal{G}_{k},\;\;k\in\{11,12,13\}$:} For the channel with $\mathcal{G}=\mathcal{G}_{11}$, we choose the first $\min\{r_3,d_1\}$ columns of $\mathbf{V}_1$ from the columns of $\mathbf{S}_3$, and randomly generate the remaining $(d_1-r_3)^+$ columns; we construct $\mathbf{V}_2$ and $\mathbf{V}_3$ the same as the ones for the channel with $\mathcal{G}=\mathcal{G}_8$. 

For the channel with $\mathcal{G}=\mathcal{G}_{12}$, we choose the first $\min\{r_2,d_1\}$ columns of $\mathbf{V}_1$ from the columns of $\mathbf{S}_2$, and randomly generate the remaining $(d_1-r_2)^+$ columns; we randomly generate the $d_2$ columns of $\mathbf{V}_2$; we choose the first $\min\{r_1,d_3\}$ columns of $\mathbf{V}_3$ from the columns of $\mathbf{S}_1$, and randomly generate the remaining $(d_3-r_1)^+$ columns. 

For the channel with $\mathcal{G}=\mathcal{G}_{13}$, we randomly generate $d_1$ columns of $\mathbf{V}_1$, and we construct $\mathbf{V}_2$ and $\mathbf{V}_3$ the same as the ones for the channel with $\mathcal{G}=\mathcal{G}_8$.

Using these matrices, we now prove achievability for the channel with $\mathcal{G}=\mathcal{G}_{11}$. We can similarly prove achievability for the channel with $\mathcal{G}=\mathcal{G}_{k},\;k\in\{12,13\}$. 

For the channel with $\mathcal{G}=\mathcal{G}_{11}$, we can almost surely have $d_1$ interference-free dimensions at receiver~1 if
\begin{align*}
d_1+(\max\{d_2,d_3\}-r_1)^+&\leq \min\{N_0,N_1\};
\end{align*}
$d_2$ interference-free dimensions at receiver~2 if
\begin{align*}
d_2&\leq \min\{N_0,N_2\};
\end{align*}
$d_3$ interference-free dimensions at receiver~3 if
\begin{align*}
d_3+(d_1-r_3)^+&\leq \min\{N_0,N_3\}.
\end{align*}

\underbar{$\mathcal{G}=\mathcal{G}_{k},\;\;k\in\{14,15\}$:}
We just need to prove achievability for the channel with 
$\mathcal{G}=\mathcal{G}_{14}$ as each $K_i$ for $\mathcal{G}_{14}$ is a subset of the corresponding one for $\mathcal{G}_{15}$.

We construct $\mathbf{V}_1$ by choosing its first $\min\{r_3,d_1\}$ columns from the columns of $\mathbf{S}_3$, and randomly generating the remaining $(d_1-r_3)^+$ columns. We randomly generate the $d_2$ columns of $\mathbf{V}_2$.  We construct $\mathbf{V}_3$ by choosing its first $\min\{r_1,d_3\}$ columns from the columns of $\mathbf{S}_1$, and randomly generating the remaining $(d_3-r_1)^+$ columns.

Using these matrices, we can almost surely have $d_1$ interference-free dimensions at receiver~1 if
\begin{align*}
d_1+(d_3-r_1)^+&\leq \min\{N_0,N_1\};
\end{align*}
$d_2$ interference-free dimensions at receiver~2 if
\begin{align*}
d_2&\leq \min\{N_0,N_2\};
\end{align*}
$d_3$ interference-free dimensions at receiver~3 if
\begin{align*}
d_3+(d_1-r_3)^+&\leq \min\{N_0,N_3\}.
\end{align*}

\underbar{$\mathcal{G}_{16}$:} We randomly generate the matrices $\{\mathbf{V}_{i}\}_{i=1}^{3}$. Since each receiver knows a priori all the requested messages by the other receivers, the channel is equivalent to three Gaussian MIMO point-to-point channel. Therefore, $(d_1,d_2,d_3)$ is achievable if it satisfies
\begin{align*}
d_i\leq\min\{N_0,N_i\},\;i\in\mathcal{V}_\mathcal{G}.
\end{align*}

\section{Achieving fractional DoF Points}\label{Sec:FractionalDoF}
In this section, we prove the achievability of all the fractional points  within the region $\mathcal{D}_k$, $k\in\{1,2,\ldots,16\}$, for the channel with $\mathcal{G}=\mathcal{G}_k$.

We here show that all the corner points of the polyhedron $\mathcal{D}_k,\;k\neq 7$, are integer points. Consequently, we can achieve the whole region using time sharing among the integer points. 

Any corner point of the polyhedron $\mathcal{D}_k,\;k\in\{1,2,\ldots, 6\}$,  is the intersection of three of the seven planes $d_1+d_2+d_3=N_0$, $d_i=N_i,\;i\in\mathcal{V}_\mathcal{G}$, and $d_i=0,\;i\in\mathcal{V}_\mathcal{G}$.
If a corner point lies on three of the last six planes, it clearly has three integer elements. If a corner point lies on $d_1+d_2+d_3=N_0$ and two of the last six planes, it also has three integer elements as having two integer elements yields the third to be an integer as well. This shows that all the corners points are integer points. 

Considering the region $\mathcal{D}_k,\;\;k\in\{8,9,10\}$, if the inequality $N_2+N_3\leq N_0$ holds, the conditions $d_1+d_2\leq N_0$, and $d_1+d_3\leq N_0$ are redundant, and the resulting polyhedron is the same as $\mathcal{D}_1$ (which we showed that all of its corner points are integer points). If $N_0<N_2+N_3$, any corner point is the intersection of three of the nine planes $d_1+d_2=N_0$, $d_1+d_3=N_0$, $d_1+d_2+d_3=N_2+N_3$, $d_i=N_i,\;i\in\mathcal{V}_\mathcal{G}$, and $d_i=0,\;i\in\mathcal{V}_\mathcal{G}$. If a corner point lies on at least one of the last six planes, having one integer element yields the other two to be integers as well. The intersection of the first three planes is 
\begin{align*}
(d_1,d_2,d_3)\hskip-2pt=\hskip-2pt(2N_0\hskip-2pt-\hskip-2ptN_2\hskip-2pt-\hskip-2ptN_3,N_2+N_3\hskip-2pt-\hskip-2ptN_0,N_2+N_3\hskip-2pt-\hskip-2ptN_0)
\end{align*} 
which is an integer point as well. This shows that all the corner points are integer points.

A similar discussion shows that all the corner points of the polyhedron $\mathcal{D}_k,\;\;k\in\{11,12,\ldots,16\}$ are integer points.

In the rest of this section, we first show that if $N_0$ is odd, and $\frac{N_0}{2}\leq N_i,\;i\in\mathcal{V}_\mathcal{G}$, the polyhedron $\mathcal{D}_7$ has one fractional corner point. Otherwise all the corner points are integer points. We then prove the achievability of the fractional corner point using two-symbol extension of our scheme in Section~\ref{Sec:IntegerDoF}. Consequently, we can achieve the whole region using time sharing.

\begin{figure}[t]
	\centering
	\includegraphics[width=0.5\textwidth]{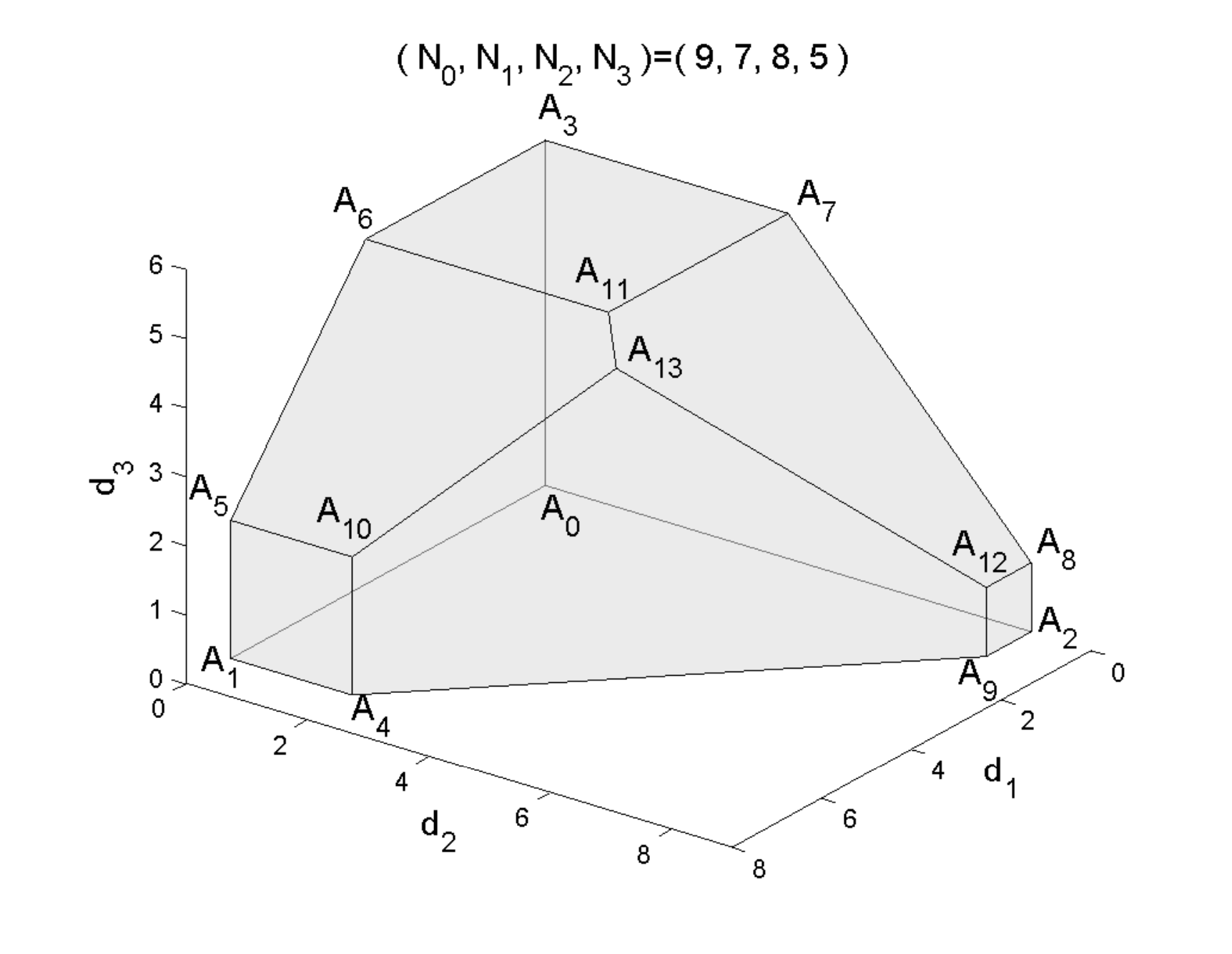}
	\vskip-0pt
	\caption{The region $\mathcal{D}_7$ for the antenna configuration $(N_0,N_1,N_2,N_3)=(9,7,8,5)$. The corner points of $\mathcal{D}_7$  are $\{A_k\}_{k=0}^{13}$ where $A_{13}=(4.5,4.5,4.5)$ is the only fractional corner point. The corner points of the DoF region without RMSI, i.e., $\mathcal{D}_1$, are $\{A_k\}_{k=0}^{9}$. This shows the improvement in the DoF region using RMSI.} 
	\vskip-0pt
	\label{Fig:OuterBoundDoFG7}
\end{figure}

Any corner point of the polyhedron $\mathcal{D}_7$ is the intersection of three of the nine planes $d_1+d_2=N_0$, $d_1+d_3=N_0$, $d_2+d_3=N_0$, $d_i=N_i,\;i\in\mathcal{V}_\mathcal{G}$, and $d_i=0,\;i\in\mathcal{V}_\mathcal{G}$. If a corner point lies on at least one of the last six planes, having one integer element yields the other two to be integers as well. The intersection of the first three planes is $(d_1,d_2,d_3)=(\frac{N_0}{2},\frac{N_0}{2},\frac{N_0}{2})$ which is a fractional corner point when $N_0$ is odd, and $\frac{N_0}{2}\leq N_i,\;i\in\mathcal{V}_\mathcal{G}$. Fig.~\ref{Fig:OuterBoundDoFG7} shows the region $\mathcal{D}_7$ and its corner points for a specific antenna configuration.

\subsection{Two-Symbol Extension}
We here prove the achievability of the DoF point $(\frac{N_0}{2},\frac{N_0}{2},\frac{N_0}{2})$ when $N_0$ is odd and $\frac{N_0}{2}\leq N_i,\;i\in\mathcal{V}_\mathcal{G}$. To achieve this point, we use our proposed scheme in Section~\ref{Sec:IntegerDoF} in conjunction with two-symbol extension over time~\cite{DoF2PairXJafar} where the channel matrices are 
\begin{align*}
\mathbf{\Theta}_i=\begin{bmatrix}
\mathbf{H}_i&\mathbf{0} \\ 
\mathbf{0}& \mathbf{H}_i
\end{bmatrix},\;i\in\mathcal{V}_\mathcal{G},
\end{align*}
$\mathbf{\Theta}_i\in\mathbb{C}^{2N_i\times2N_0}$. We again construct three subcodebooks. Subcodebook~$i$, $i\in\mathcal{V}_\mathcal{G}$, consists of $2^{nR_i}$ i.i.d. codewords 
\begin{align*}
\mathbf{X}'^{\frac{n}{2}}_i(m_i)=(\mathbf{X}'_{i,1}(m_i),\mathbf{X}'_{i,2}(m_i),\ldots,\mathbf{X}'_{i,\frac{n}{2}}(m_i)),
\end{align*}
generated according to $\prod_{j=1}^{\frac{n}{2}}p_{{\mathbf{X}^{'}_i}}(\mathbf{x}'_{i,j})$ where $\mathbf{X}'_i\in\mathbb{C}^{N_0\times 1}$, $\mathbf{X}'_i\sim\mathcal{CN}(0,\mathbf{\Sigma}'_i)$, $\mathbf{\Sigma}'_i$ is a $N_0\times N_0$ diagonal matrix, $\text{tr}(\mathbf{\Sigma}'_i)=2P_i$, and $\sum_{i=1}^{3}P_i=P$. We then construct the transmitted codeword as
\begin{align*}
\begin{bmatrix}
\mathbf{X}_{0,2j-1}(m_1,m_2,m_3)\\ 
\mathbf{X}_{0,2j}(m_1,m_2,m_3)
\end{bmatrix}\hskip-4pt=\hskip-3pt\sum_{i=1}^{3}\mathbf{U}_{i}\mathbf{X}^{'}_{i,j}(m_i),\;j\hskip-3pt\in\hskip-3pt\{1,2,\ldots,\frac{n}{2}\},
\end{align*}
where $\mathbf{U}_{i}\in\mathbb{C}^{2N_0\times N_0}$ has columns, each of Euclidean norm one. We choose the first $\min\{2r_q,N_0\}$ columns of $\mathbf{U}_{i}$, $i\in\mathcal{V}_\mathcal{G}$, from the columns of $\mathbf{T}_q$, $q=(i\hskip-4pt\mod 3)+1$, which is randomly generated in the null space of $\mathbf{\Theta}_q$, i.e., $\mathbf{\Theta}_q\mathbf{T}_q=0$. We randomly generate the remaining $(N_0-2r_q)^+$ columns of $\mathbf{U}_i$ according to an isotropic distribution. Using this construction, the matrices $[\mathbf{U}_1,\mathbf{U}_3]$, $[\mathbf{U}_2,\mathbf{U}_1]$, and $[\mathbf{U}_3,\mathbf{U}_2]$ are almost surely full rank.

We perform decoding every two symbols which provides us with $2N_i$ space dimensions at receiver~$i$. Since $\frac{N_0}{2}\leq N_i$, the inequality
\begin{align*}
N_0+(N_0-2r_i)^+&\leq \min\{2N_0,2N_i\},\;i\in\mathcal{V}_\mathcal{G},
\end{align*}
holds, which yields the non-zero columns of the matrix $\left[\mathbf{\Theta}_i\mathbf{U}_i,\mathbf{\Theta}_i\mathbf{U}_q\right],\;i\in\mathcal{V}_\mathcal{G},\,q=((i+1)\hskip-4pt\mod 3)+1$, to be almost surely linearly independent. Then, at each receiver, the number of interference-free dimensions over two symbols is $N_0$, and we have achieved the DoF point $(\frac{N_0}{2},\frac{N_0}{2},\frac{N_0}{2})$.

\section{Outer Bound on the DoF Region}\label{Sec:ConverseDoF}
In this section, we first prove two lemmas. We then present the converse proof for Theorem~\ref{Theorem:DoF} using these two lemmas.
\begin{lemma}\label{Lemma:OuterusingIAS}
If $(d_1,d_2,d_3)$ is achievable for the three-receiver Gaussian MIMO broadcast channel with RMSI, then it must satisfy
\begin{align*}
\sum_{k\in\mathcal{V}_\mathcal{Q}}d_k&\leq \min\{N_0,\sum_{k\in\mathcal{V}_\mathcal{Q}}N_k\},
\end{align*}
for every acyclic induced subgraph $\mathcal{Q}$ of the side information graph ($\mathcal{V}_\mathcal{Q}$ is the vertex set of $\mathcal{Q}$).
\end{lemma}
\begin{IEEEproof}
	The proof is presented in Appendix~A.	
\end{IEEEproof}

\begin{lemma}\label{Lemma:3MCCapacity}
	Considering the three-receiver memoryless broadcast channel with RMSI where channel input is $X_0$, channel outputs are $Y_i,\;i\in\mathcal{V}_\mathcal{G}$, $K_1\subseteq\{M_2,M_3\}$, $K_2=\{M_3\}$,  $K_3=\{M_2\}$, and $X_0\rightarrow Y_1\rightarrow (Y_2,Y_3)$ form a Markov chain, the capacity region is the closure of the set of all rate triples $(R_1,R_2,R_3)$, each satisfying  
	\begin{align*}
	R_1&<I(X_0;Y_1\mid U_0),\\
	R_2&<I(U_0;Y_2),\\
	R_3&<I(U_0;Y_3),
	\end{align*}
	for some distribution $p(u_0,x_0)$.
\end{lemma}
\begin{IEEEproof}
	The proof is presented in Appendix B.	
\end{IEEEproof}

We here present the converse proof for Theorem~\ref{Theorem:DoF}.

\underbar{$\mathcal{G}=\mathcal{G}_k,\;k\in\{1,2,\ldots, 16\}\setminus\{8,9,10\}$:} Lemma~\ref{Lemma:OuterusingIAS} provides a tight outer bound for all these side information configurations.

\underbar{$\mathcal{G}=\mathcal{G}_k,\;k\in\{8,9,10\}$:} Using Lemma~\ref{Lemma:OuterusingIAS}, we obtain the necessary conditions
\begin{align*}
d_1+d_2&\leq N_0,\\
d_1+d_3&\leq N_0,\\
d_i&\leq N_i,\;i\in\mathcal{V}_\mathcal{G}.
\end{align*}
If at least one of the three inequalities $N_2\geq N_0$, $N_3\geq N_0$, or $N_0\geq N_1+N_2+N_3$ holds, the condition 
\begin{align}
d_1+d_2+d_3\leq\max\{N_0,N_2+N_3\}\label{specialcondition},
\end{align}
in the achievable region is redundant, and the converse proof for these side information configurations is complete. Otherwise, (i.e., when $N_2<N_0$, $N_3<N_0$, and $N_0<N_1+N_2+N_3$), to show that the condition in \eqref{specialcondition} is also a necessary condition, we construct an enhanced channel by providing the channel outputs at receivers~2 and~3 to receiver~1. In the enhanced channel, the channel output at receiver~1 is $(\mathbf{Y}_1,\mathbf{Y}_2,\mathbf{Y}_3)$. Since $\mathbf{X}_0\rightarrow(\mathbf{Y}_1,\mathbf{Y}_2,\mathbf{Y}_3)\rightarrow(\mathbf{Y}_2,\mathbf{Y}_3)$ form a Markov chain, we use Lemma~\ref{Lemma:3MCCapacity} to bound the sum-rate as follows.
\begin{align}
\hskip-0.5ptR_1\hskip-2pt+\hskip-2ptR_2\hskip-2pt+\hskip-2ptR_3\hskip-2pt&\leq \hskip-2ptI(\mathbf{X}_0;\hskip-2pt\mathbf{Y}_1,\mathbf{Y}_2,\mathbf{Y}_3\hskip-2pt\mid\hskip-2pt U_0)\hskip-2pt+\hskip-2ptI(U_0;\hskip-2pt\mathbf{Y}_2)\hskip-2pt+\hskip-2ptI(U_0;\hskip-2pt\mathbf{Y}_3)\nonumber\\
&=h(\mathbf{Y}_1,\mathbf{Y}_2,\mathbf{Y}_3\mid U_0)-h(\mathbf{Z}_1,\mathbf{Z}_2,\mathbf{Z}_3)\nonumber\\
&\hskip9pt+h(\mathbf{Y}_2)-h(\mathbf{Y}_2\mid U_0)+h(\mathbf{Y}_3)\hskip-2pt-\hskip-2pth(\mathbf{Y}_3\mid U_0)\nonumber\\
&\leq h(\mathbf{Y}_1\mid \mathbf{Y}_2,\mathbf{Y}_3, U_0)
\hskip-2pt+h(\mathbf{Y}_2)\hskip-2pt+h(\mathbf{Y}_3)\nonumber\\
&\hskip130pt-h(\mathbf{Z}_1,\mathbf{Z}_2,\mathbf{Z}_3)\nonumber\\
&= h(\mathbf{Y}_1\mid \mathbf{Y}_2,\mathbf{Y}_3, U_0)
+h(\mathbf{Y}_2)+h(\mathbf{Y}_3)\nonumber\\
&\hskip95pt-h(\mathbf{Z}_1)\hskip-2pt-\hskip-2pth(\mathbf{Z}_2)\hskip-2pt-\hskip-2pth(\mathbf{Z}_3)\nonumber\\
&=I(\mathbf{X}_0;\mathbf{Y}_2)+I(\mathbf{X}_0;\mathbf{Y}_3)\nonumber\\
&\hskip57pt+h(\mathbf{Y}_1\mid \mathbf{Y}_2,\mathbf{Y}_3, U_0)-h(\mathbf{Z}_1)\label{DoFOuterG81}.
\end{align}
From the Gaussian MIMO point-to-point channel, the mutual information terms in~\eqref{DoFOuterG81} are upper bounded as~\cite{MIMOBook}
\begin{align}
I(\mathbf{X}_0;\mathbf{Y}_2) \leq N_2\log P+o(\log P),\label{DoFOuterG82}\\
I(\mathbf{X}_0;\mathbf{Y}_3) \leq N_3\log P+o(\log P),\label{DoFOuterG83}
\end{align}
where $\underset{P\rightarrow\infty}{\lim}\frac{o(\log P)}{\log{P}}\rightarrow 0$. In order to upper bound the remaining terms in~\eqref{DoFOuterG81}, i.e.,  
\begin{align*}
h(\mathbf{Y}_1\mid \mathbf{Y}_2,\mathbf{Y}_3, U_0)-h(\mathbf{Z}_1),
\end{align*}
we generalize the technique used by Weingarten et al. for a two-receiver compound broadcast channel~\cite[Theorem 4]{DoFCompoundBC}. We consider two cases: \textit{Case}~I where $N_0\leqslant N_2+N_3$, and \textit{Case}~II where $N_2+N_3<N_0$.

For \textit{Case}~I ($N_2,N_3<N_0\leqslant N_2+N_3$), we define $\mathbf{Y}'$, $\mathbf{H}'$, and $\mathbf{Z}'$ as
\begin{align*}
\underset{\mathbf{Y}'}{\underbrace{\begin{bmatrix}
\mathbf{Y}_{2}\\ 
Y_{3[1]}\\ 
\vdots\\ 
Y_{3[N_0-N_2]}
\end{bmatrix}}}=\underset{\mathbf{H}'}{\underbrace{\begin{bmatrix}
\mathbf{H}_{2}\\ 
\mathbf{H}_{3[1:]}\\ 
\vdots\\ 
\mathbf{H}_{3[N_0-N_2:]}
\end{bmatrix}}}\mathbf{X}_0+\underset{\mathbf{Z}'}{\underbrace{\begin{bmatrix}
\mathbf{Z}_{2}\\ 
Z_{3[1]}\\ 
\vdots\\ 
Z_{3[N_0-N_2]}
\end{bmatrix}}}.
\end{align*}
Since $\mathbf{H}'\in\mathbb{C}^{N_0\times N_0}$ is almost surely full rank, $\mathbf{H}_1$ can be written as
$\mathbf{H}_{1}=\mathbf{\Lambda}'\mathbf{H}'$ where $\mathbf{\Lambda}'\in\mathbb{C}^{N_1\times N_0}$. Then we have
\begin{align*}
\mathbf{Y}_1&=\mathbf{\Lambda}'\mathbf{H}'\mathbf{X}_0+\mathbf{Z}_1\\
&=\mathbf{\Lambda}'(\mathbf{Y}'-\mathbf{Z}')+\mathbf{Z}_{1},
\end{align*}
which results in
\begin{align}
&h(\mathbf{Y}_1\mid \mathbf{Y}_2,\mathbf{Y}_3, U_0)-h(\mathbf{Z}_1)\nonumber\\
&\hskip20pt=h(-\mathbf{\Lambda}'\mathbf{Z}'+\mathbf{Z}_{1}\mid \mathbf{Y}_2,\mathbf{Y}_3,U_0)-h(\mathbf{Z}_1)\nonumber\\
&\hskip20pt\leq h(-\mathbf{\Lambda}'\mathbf{Z}'+\mathbf{Z}_{1})-h(\mathbf{Z}_1)\nonumber\\
&\hskip20pt=o(\log P)\label{DoFOuterG84}.
\end{align}
Using \eqref{DoFOuterG81}--\eqref{DoFOuterG84}, the converse proof for \textit{Case}~I is complete.

For \textit{Case}~II ($N_2+N_3<N_0<N_1+N_2+N_3$), we define $\mathbf{Y}''$, $\mathbf{H}''$, and $\mathbf{Z}''$ as
\begin{align*}
\underset{\mathbf{Y}''}{\underbrace{\begin{bmatrix}
		Y_{1[1]}\\ 
		\vdots\\ 
		Y_{1[N_0-N_2-N_3]}\\
		\mathbf{Y}_{2}\\ 
		\mathbf{Y}_{3}
		\end{bmatrix}}}\hskip-2pt=\hskip-2pt\underset{\mathbf{H}''}{\underbrace{\begin{bmatrix}
		\mathbf{H}_{1[1:]}\\ 
		\vdots\\ 
		\mathbf{H}_{1[N_0-N_2-N_3:]}\\
		\mathbf{H}_{2}\\ 
		\mathbf{H}_{3}
		\end{bmatrix}}}\mathbf{X}_0\hskip-2pt+\hskip-4pt\underset{\mathbf{Z}''}{\underbrace{\begin{bmatrix}
		Z_{1[1]}\\ 
		\vdots\\ 
		Z_{1[N_0-N_2-N_3]}\\
		\mathbf{Z}_{2}\\
		\mathbf{Z}_{3}
		\end{bmatrix}}}.
\end{align*}
Since $\mathbf{H}''\in\mathbb{C}^{N_0\times N_0}$ is almost surely full-rank, we can write
\begin{align*}
\begin{bmatrix}
\mathbf{H}_{1[N_0-N_2-N_3+1:]}\\ 
\vdots\\ 
\mathbf{H}_{1[N_1:]}
\end{bmatrix}=\mathbf{\Lambda}''\mathbf{H}'',
\end{align*}
where $\mathbf{\Lambda}''\in\mathbb{C}^{N_1+N_2+N_3-N_0\times N_0}$. Then we have
\begin{align*}
\begin{bmatrix}
Y_{1[N_0-N_2-N_3+1]}\\ 
\vdots\\ 
Y_{1[N_1]}
\end{bmatrix}=
\mathbf{\Lambda}''\mathbf{H}''\mathbf{X}_0+
\begin{bmatrix}
Z_{1[N_0-N_2-N_3+1]}\\ 
\vdots\\ 
Z_{1[N_1]}
\end{bmatrix},
\end{align*}
where $\mathbf{H}''\mathbf{X}_0=\mathbf{Y}''-\mathbf{Z}''$. This results in 
\begin{align}
& h(\mathbf{Y}_1\mid \mathbf{Y}_2,\mathbf{Y}_3, U_0)-h(\mathbf{Z}_1)\nonumber\\
&\hskip5pt=h([Y_{1[1]},\ldots,Y_{1[N_0-N_2-N_3]}]^T\mid \mathbf{Y}_2,\mathbf{Y}_3, U_0)\nonumber\\
&\hskip15pt+h(-\mathbf{\Lambda}''\mathbf{Z}''+[Z_{1[N_0-N_2-N_3+1]},\ldots,Z_{1[N_1]}]^T\mid \mathbf{Y}'', U_0)\nonumber\\
&\hskip15pt-h(\mathbf{Z}_1)\nonumber\\
&\hskip5pt\leq h([Y_{1[1]},\ldots,Y_{1[N_0-N_2-N_3]}]^T)\nonumber\\
&\hskip15pt+h(-\mathbf{\Lambda}''\mathbf{Z}''+[Z_{1[N_0-N_2-N_3+1]},\ldots,Z_{1[N_1]}]^T)\nonumber\\
&\hskip15pt-h(\mathbf{Z}_1)\nonumber\\
&\hskip5pt=I(\mathbf{X}_0;[Y_{1[1]},\ldots,Y_{1[N_0-N_2-N_3]}]^T)\nonumber\\
&\hskip15pt+h(-\mathbf{\Lambda}''\mathbf{Z}''+[Z_{1[N_0-N_2-N_3+1]},\ldots,Z_{1[N_1]}]^T)\nonumber\\
&\hskip15pt-h([Z_{1[N_0-N_2-N_3+1]},\ldots,Z_{1[N_1]}]^T)\nonumber\\
&\hskip5pt\leq(N_0-N_2-N_3)\log P+o(\log P).\label{DoFOuterG85}
\end{align}
Using \eqref{DoFOuterG81}, \eqref{DoFOuterG82}, \eqref{DoFOuterG83}, and \eqref{DoFOuterG85}, the converse proof for \textit{Case}~II is complete.

This completes the converse proof for Theorem~\ref{Theorem:DoF}.

\section{Remarks on the Schemes and the DoF Region}
In this section, we provide some remarks on the proposed transmission schemes, and the DoF region for the three-receiver Gaussian MIMO broadcast channel with RMSI. These remarks provide some hints about the DoF region for the channel when there are four or more receivers.

\subsection{On the Transmission Schemes}
In this subsection, we make the observation that we can achieve the DoF region for all 16 possible RMSI configurations using three transmission schemes. One for the side information configurations $\{\mathcal{G}_1,\mathcal{G}_2,\ldots,\mathcal{G}_6\}$ (the scheme for the channel without RMSI, i.e., $\mathcal{G}_1$, can be used for all these configurations as they have the same DoF region), one for $\mathcal{G}_7$, and one for the rest, i.e., $\{\mathcal{G}_8,\mathcal{G}_{9},\ldots,\mathcal{G}_{16}\}$. For $\mathcal{G}_9$ and $\mathcal{G}_{10}$, the scheme for $\mathcal{G}_8$ can be used as they have the same DoF region. For $\mathcal{G}_k$, $k\in\{11,12,\ldots,16\}$, based on the following two points, the region achieved by the scheme for $\mathcal{G}_8$ is at least as large as the region achieved by the scheme for this configuration. First, the construction of $\mathbf{V}_1$ for $\mathcal{G}_k$ is a special case of the construction of $\mathbf{V}_1$ for $\mathcal{G}_8$. Second, the construction of $\mathbf{V}_2$ and $\mathbf{V}_3$ for $\mathcal{G}_k$ can be viewed as a modification of those for $\mathcal{G}_8$ in the following way: we replace some columns of these matrices that are constructed using i) the side information of the receivers, and/or ii) zero forcing in the scheme for $\mathcal{G}_8$, with the columns that are constructed independently and randomly in the scheme for $\mathcal{G}_k$. The replacement is done---depending on the extra side information in $\mathcal{G}_k$ compared to $\mathcal{G}_8$---in a way that we still have the same number of interference-free dimensions at the receivers. Consequently, the scheme for $\mathcal{G}_8$ can also achieve the DoF region of the channel with $\mathcal{G}=\mathcal{G}_k$, $k\in\{11,12,\ldots,16\}$.

\subsection{On the DoF Region with RMSI}
In this subsection, we introduce some properties of the DoF region of the Gaussian MIMO broadcast channel with RMSI under two specific antenna configurations where i) the number of antennas at the transmitter is greater than or equal to the sum-number of antennas at the receivers, and ii) the number of antennas at all the nodes are equal.

\subsubsection{$N_0\geq N_1+N_2+N_3$} Under this configuration, according to Theorem~\ref{Theorem:DoF}, the DoF region of the channel is
\begin{align*}
\mathcal{D}_{k}=\Big{\{}(d_1,d_2,d_3)\in&\mathbb{R}_+^3 \mid
d_i\leq N_i,\;i\in\mathcal{V}_\mathcal{G}\Big{\}},\;\forall k.
\end{align*}
This shows that the side information available at the receivers cannot enlarge the DoF region of the channel when $N_0\geq N_1+N_2+N_3$. This is because zero forcing can cancel the interference at all the receivers, and the transmitter can simultaneously create three independent virtual MIMO point-to-point channels, one to each receiver~$i$ with DoF $N_i$. Then the side information which is used to alleviate the interference at the receivers is no longer useful as far as the DoF region is concerned.

\subsubsection{$N_0\hskip-2pt=\hskip-2ptN_1\hskip-2pt=\hskip-2ptN_2\hskip-2pt=\hskip-2ptN_3$} Under this configuration, according to Theorem~\ref{Theorem:DoF}, the DoF region of the channel is
\begin{align}
\mathcal{D}_{k}=\Big{\{}(d_1,d_2,d_3)\in&\mathbb{R}_+^3 \mid
\sum_{k\in\mathcal{V}_\mathcal{Q}}d_k\leq N_0,\;\forall\mathcal{Q}\Big{\}},\;\forall k,\label{DoFEqualAntenna}
\end{align}
where $\mathcal{Q}$ is an acyclic induced subgraph of the side information graph of the channel. 

In this configuration, as opposed to the previous configuration, we cannot perform zero forcing, and the side information plays a key role. Based on this, and as the DoF region is defined in the high signal-to-noise ratio region, we show that there are some common properties between the DoF region and the capacity region of the index coding problem~\cite{IndexCoding}. The three-receiver index coding problem considers a noiseless broadcast channel with RMSI where there are three messages, $M_i\in\mathcal{M}_i$, $i\in\mathcal{V}_\mathcal{G}$, each requested by one receiver, and a common noiseless link that carries $n$ bits. Arbabjolfaei et al.~\cite{CapacityRegionIndexCoding1} established the capacity region of the index coding problem for up to five receivers. We present their result for the three-receiver case as Proposition~\ref{Proposition:CapIndexCoding}. The capacity region of the three-receiver index coding problem is achieved using flat coding and time sharing~{\cite[Section III]{CapacityRegionIndexCoding1}}.

\begin{proposition}\label{Proposition:CapIndexCoding}
The capacity region of the three-receiver index coding problem is the set of all rate triples $(R_1,R_2,R_3)$, each satisfying 
\begin{align}
\sum_{k\in\mathcal{V}_\mathcal{Q}}R_k\leq 1,\; \forall \mathcal{Q},\label{CapIndexCoding}
\end{align}
where $\mathcal{Q}$ is an acyclic induced subgraph of the side information graph of the channel.
\end{proposition}

The first property that we can see from~\eqref{DoFEqualAntenna} and~\eqref{CapIndexCoding} is that the DoF region normalized by the number of antennas at each node is the same as the capacity region of index coding.

Another property for the capacity region of index coding is that removing the arcs that are not part of a directed cycle does not decrease the capacity region~\cite{CapacityRegionIndexCoding2}. According to~\eqref{DoFEqualAntenna}, this property is also valid for the DoF region of the three-receiver Gaussian MIMO broadcast channel with the same number of antennas at all the nodes.

However, this property is not valid for the DoF region in general. For instance, the DoF region of the channel with $\mathcal{G}=\mathcal{G}_{11}$ is strictly larger than the one of the channel with $\mathcal{G}=\mathcal{G}_{8}$ when $N_2<N_0$, $N_3<N_0$, and $N_0<N_1+N_2+N_3$. This property is not valid also for the capacity region of the Gaussian MIMO broadcast channel when even all the nodes have the same number of antennas. For example, in a previous study~\cite{ThreeReceiverAWGNwithMSI}, we established the capacity region of the Gaussian channel with the side information graphs $\mathcal{G}_{8}$ and $\mathcal{G}_{11}$ where all the nodes have one antenna. The results show that the capacity region of the channel with $\mathcal{G}=\mathcal{G}_{11}$ is strictly larger than the one of the channel with $\mathcal{G}=\mathcal{G}_{8}$ when the absolute value of the channel gain for receiver~1 is the largest one and the one for receiver~3 is the smallest one.

\section{Conclusion}
We considered the three-receiver Gaussian multiple-input multiple-output (MIMO) broadcast channel with an arbitrary number of antennas at each node. We assumed that (i) channel matrices are known to all the nodes, (ii) the receivers have private-message requests, and (iii) each receiver may know some of the messages requested by the other receivers as receiver message side information (RMSI). We established the degrees-of-freedom (DoF) region of the channel for all 16 possible non-isomorphic RMSI configurations. To this end, we first proposed a scheme for each RMSI configuration which utilizes both the null space and the side information of the receivers. We used our schemes in conjunction with time sharing for 15 RMSI configurations, and with time sharing and two-symbol extension for the remaining one. We then derived a tight outer bound for each RMSI configuration by constructing enhanced versions of the channel, and upper bounding their DoF region. Furthermore, we showed that some properties for the capacity region of the index coding problem also hold for the DoF region where all the nodes have the same number of antennas.

\section*{Appendix A}\label{AppendixA}
In this section, we present the proof of Lemma~\ref{Lemma:OuterusingIAS}.
\begin{IEEEproof}
Any acyclic induced subgraph, $\mathcal{Q}$, of a side information graph represents a channel with RMSI where there are three or fewer receivers. We construct an enhanced channel for this channel. To this end, we first choose a receiver of this channel with outdegree zero, say receiver~$\ell$, $\ell\in\mathcal{V}_\mathcal{Q}$ (outdegree of a vertex (receiver) is the number of its outgoing arcs). We then provide the channel outputs at the other receivers of this channel to receiver~$\ell$. In the enhanced channel, receiver~$\ell$ can first decode its own message and the messages of the other receivers with outdgeree zero (if any) as it has all the information using which they decode their messages. Receiver~$\ell$ can then decode the messages of the receivers whose side information is already decoded at this receiver. By continuing this approach, as we have an acyclic subgraph, receiver~$\ell$ can decode all the messages $\{M_k\}$, $k\in\mathcal{V}_\mathcal{Q}$. Consequently, from the Gaussian MIMO point-to-point channel where a transmitter with $N_0$ antennas wants to transmit the messages $\{M_k\}$, $k\in\mathcal{V}_\mathcal{Q}$, to a receiver with $\sum_{k\in\mathcal{V}_\mathcal{Q}}N_k$ antennas, we obtain the necessary condition
\begin{align*}
\sum_{k\in\mathcal{V}_\mathcal{Q}}d_k&\leq \min\{N_0,\sum_{k\in\mathcal{V}_\mathcal{Q}}N_k\}.
\end{align*}
\vskip-35pt
\end{IEEEproof}
\vskip30pt
\section*{Appendix B}\label{AppendixB}
In this section, we present the proof of Lemma~\ref{Lemma:3MCCapacity}.
\begin{IEEEproof}
	(\textit{Achievability}) To construct the codebook, we first convert the messages $M_2$ and $M_3$ into binary vectors, and XOR them, i.e., $M_\text{x}=M_2\oplus M_3$,  where $\oplus$ denotes the bitwise XOR operation with zero padding for messages of unequal length ($M_\text{x}$ is an $n\max\{R_2,R_3\}$-bit message). We then, choose a distribution $p_{_{U_0,X_0}}(u_0,x_0)$, and generate $2^{n\max\{R_2,R_3\}}$ codewords 
	\begin{align*}
	U_0^n(m_\text{x})=(U_{0,1}(m_\text{x}),U_{0,2}(m_\text{x}),\ldots,U_{0,n}(m_\text{x})),
	\end{align*}
	according to $\prod_{j=1}^np_{_{U_0}}(u_{0,j})$. We finally, using superposition coding, generate $2^{nR_1}$ codewords $X_0^n(m_\text{x}, m_1)$ for each $U_0^n(m_\text{x})$ according to $\prod_{j=1}^np_{_{X_0\mid U_0}}(x_{0,j}\mid u_{0,j}(m_\text{x}))$.
	
	Receiver~1 decodes $\hat{m}_1$ if there exits a unique $\hat{m}_1$ such that $\left(U_0^n(m_\text{x}),X_0^n(m_\text{x},\hat{m}_1),Y_1^n\right)\in\mathcal{T}_{\epsilon}^{n}$ for some $m_\text{x}$ where $\mathcal{T}_\epsilon^n$ is the set of jointly $\epsilon$-typical $n$-sequences with respect to the considered distribution~\cite[p. 30]{NITBook}; otherwise an error is declared. We assume without loss of generality that the transmitted messages are equal to zero by the symmetry of the codebook construction. Then receiver~1 makes a decoding error only if one or more of the following events occur.
	\begin{align*}
	\mathcal{E}_{11}&:\left(U_0^n(0),X_0^n(0,0),Y_1^n\right)\notin\mathcal{T}_{\epsilon}^{n},\\
	\mathcal{E}_{12}&:\left(U_0^n(0),X_0^n(0,m_1),Y_1^n\right)\in\mathcal{T}_{\epsilon}^{n}\;\;\text{for some }m_1\neq 0,\\
	\mathcal{E}_{13}&:\left(U_0^n(m_\text{x}),X_0^n(m_\text{x},m_1),Y_1^n\right)\in\mathcal{T}_{\epsilon}^{n}\\
	&\hskip120pt\text{for some }m_\text{x}\neq 0, m_1\neq 0.
	\end{align*}
	According to these error events, and using the packing lemma~\cite[p. 45]{NITBook}, receiver~1 can reliably decode $M_1$ if
	\begin{align}
	R_1&<I(X_0;Y_1\mid U_0),\label{achiev11}\\
	R_1+\max\{R_2,R_3\}&<I(X_0;Y_1).\label{achiev12}
	\end{align}
	
	Since receiver~2 knows $m_3$ a priori, it decodes $\hat{m}_2$ if there exits a unique $\hat{m}_2$ such that $\left(U_0^n(\hat{m}_2\oplus 0),Y_2^n\right)\in\mathcal{T}_{\epsilon}^{n}$; otherwise an error is declared. Then receiver~2 makes a decoding error only if one or more of the following events occur.
	\begin{align*}
	\mathcal{E}_{21}&:\left(U_0^n({0}),Y_2^n\right)\notin\mathcal{T}_{\epsilon}^{n},\\
	\mathcal{E}_{22}&:\left(U_0^n(m_2\oplus {0}),Y_2^n\right)\in\mathcal{T}_{\epsilon}^{n}\text{ for some }m_2\neq{0}.
	\end{align*}
	Hence, using the packing lemma, receiver~2 can reliably decode $M_2$ if $R_2<I(U_0;Y_2)$.
	
	Since receiver~3 knows $m_2$ a priori, it decodes $\hat{m}_3$ if there exits a unique $\hat{m}_3$ such that $\left(U_0^n(0\oplus \hat{m}_3),Y_3^n\right)\in\mathcal{T}_{\epsilon}^{n}$; otherwise an error is declared. Then receiver~3 makes a decoding error only if one or more of the following events occur.
	\begin{align*}
	\mathcal{E}_{31}&:\left(U_0^n({0}),Y_3^n\right)\notin\mathcal{T}_{\epsilon}^{n},\\
	\mathcal{E}_{32}&:\left(U_0^n({0}\oplus m_3 ),Y_3^n\right)\in\mathcal{T}_{\epsilon}^{n}\text{ for some }m_3\neq\mathbf{0}.
	\end{align*}
	Hence, using the packing lemma, receiver~3 can reliably decode $M_3$ if $R_3<I(U_0;Y_3)$.
	
	Since $U_0\rightarrow X_0\rightarrow Y_1\rightarrow (Y_2,Y_3)$ form a Markov chain, we have $I(U_0;Y_i)\leq I(U_0;Y_1),\;i=2,3$. This makes the condtion in \eqref{achiev12} redundant, and completes the achievability proof. Note that receiver~1 does not use its side information during the decoding process. Then the achievability proof is irrespective of $K_1$. 
	
	(\textit{Converse}) By Fano's inequality~\cite[p. 19]{NITBook}, we have
	\begin{align}
	H(M_1\mid Y_1^n,M_2,M_3)&\leq n\epsilon_n,\label{fano1111}\\
	H(M_2\mid Y_2^n,M_3)&\leq n\epsilon_n,\label{fano1112}\\
	H(M_3\mid Y_3^n,M_2)&\leq n\epsilon_n,\label{fano1113}
	\end{align}
	where $\epsilon_n\rightarrow 0$ as $n\rightarrow \infty$. Using \eqref{fano1111}--\eqref{fano1113}, if a rate triple is achievable, then it must satisfy
	\begin{align}
		nR_1&\leq I(M_1;Y_1^n\mid M_2,M_3)+n\epsilon_n,\label{conv1111}\\
	    nR_2&\leq I(M_2;Y_2^n\mid M_3)+n\epsilon_n,\label{conv1112}\\
	    nR_3&\leq I(M_3;Y_3^n\mid M_2)+n\epsilon_n.\label{conv1113}
	\end{align}
	We define the auxiliary random variable $U_{0,j}=(M_2,M_3,Y_1^{j-1})$, where $Y_1^{j-1}=(Y_{1,1},Y_{1,2},\ldots,Y_{1,j-1})$, and expand the mutual information terms in \eqref{conv1111}--\eqref{conv1113} respectively as follows.
	\begin{align*}
	nR_1&\leq I(M_1;Y_1^n\mid M_2,M_3)+n\epsilon_n\\
	&=\sum_{j=1}^{n}I(M_1;Y_{1,j}\mid Y_1^{j-1},M_2,M_3)+n\epsilon_n\\
	&\overset{(a)}{=}\sum_{j=1}^{n}I\left(X_{0,j};Y_{1,j}\mid Y_1^{j-1},M_2,M_3\right)+n\epsilon_n\\
	&=\sum_{j=1}^{n}I\left(X_{0,j};Y_{1,j}\mid U_{0,j}\right)+n\epsilon_n,
	\end{align*}
	\begin{align*}
	nR_2&\leq I(M_2;Y_2^n\mid M_3)+n\epsilon_n\\
	&=\sum_{j=1}^{n}I(M_2;Y_{2,j}\mid Y_2^{j-1},M_3)+n\epsilon_n\\
	&\leq\sum_{j=1}^{n}I(M_2,M_3,Y_2^{j-1};Y_{2,j})+n\epsilon_n\\
	&\leq\sum_{j=1}^{n}I(M_2,M_3,Y_2^{j-1},Y_1^{j-1};Y_{2,j})+n\epsilon_n\\
	&\overset{(b)}{=}\sum_{j=1}^{n}I(M_2,M_3,Y_1^{j-1};Y_{2,j})+n\epsilon_n\\
	&=\sum_{j=1}^{n}I(U_{0,j};Y_{2,j})+n\epsilon_n,	
	\end{align*}
	and
	\begin{align*}
	nR_3&\leq I(M_3;Y_3^n\mid M_2)+n\epsilon_n\\
	&=\sum_{j=1}^{n}I(M_3;Y_3(j)\mid Y_3^{j-1},M_2)+n\epsilon_n\\
	&\leq\sum_{j=1}^{n}I(M_2,M_3,Y_3^{j-1};Y_{3,j})+n\epsilon_n\\
	&\leq\sum_{j=1}^{n}I(M_2,M_3,Y_3^{j-1},Y_1^{j-1};Y_{3,j})+n\epsilon_n\\
	&\overset{(b)}{=}\sum_{j=1}^{n}I(M_2,M_3,Y_1^{j-1};Y_{3,j})+n\epsilon_n\\
	&=\sum_{j=1}^{n}I(U_{0,j};Y_{3,j})+n\epsilon_n,	
	\end{align*}
	where $(a)$ follows since $X_{0,j}$ is a function of the messages $\{M_i\}_{i=1}^{3}$, and $(M_1,M_2,M_3,Y_1^{j-1})\rightarrow X_{0,j}\rightarrow Y_{1,j}$ form a Markov chain; $(b)$ follows form the Markov chain $X_0\rightarrow Y_1\rightarrow (Y_2,Y_3)$  which implies $Y_2^{j-1}\rightarrow (M_2,M_3,Y_1^{j-1})\rightarrow Y_{2,j}$, and $Y_3^{j-1}\rightarrow (M_2,M_3,Y_1^{j-1})\rightarrow Y_{3,j}$. Since $\epsilon_n\rightarrow\infty$ as $n\rightarrow\infty$, using the standard time-sharing argument~\cite[p. 114]{NITBook} completes the converse proof. 
\end{IEEEproof}

\bibliographystyle{IEEEtran}

\end{document}